\documentclass[aps,prl,showpacs,twocolumn]{revtex4}
\usepackage{graphicx}
\usepackage{bm}
\begin{document}
\preprint{}
\title{Signal detection without finite-energy limits to quantum resolution}
\author{Alfredo Luis}
\email{alluis@fis.ucm.es}
\affiliation{Departamento de \'{O}ptica, Facultad de Ciencias
F\'{\i}sicas, Universidad Complutense, 28040 Madrid, Spain}
\date{\today}

\begin{abstract}
We show that there are extremely simple signal detection schemes where 
the finiteness of energy resources places no limit to the resolution. 
On the contrary, larger resolution can be obtained with lower energy.
To this end the generator of the signal-dependent transformation 
encoding the signal information on the probe state must be different 
from the energy. We show that the larger the deviation of the probe 
state from being minimum-uncertainty state, the better the resolution. 
\end{abstract}


\maketitle

\section{Introduction}

Precise detection provides a readily practical application of quite 
fundamental quantum ideas, such as quantum statistics, nonclassical 
states, and uncertainty relations. Quantum metrology is framed by two 
widespread beliefs: (i) for fixed mean energy, quantum uncertainty 
limits the resolution in the detection of signals with unavoidable 
bounds valid for all system states, and (ii) these bounds depend on the
energy resources employed so that better resolution requires larger 
amounts of energy. This is the case of the well-known standard quantum 
limit and Heisenberg limit \cite{HL,D0,D1,D2,D3,D4,D5,D6,D7,NL1,NL2}. 

Here we show that there are practical schemes where energy finiteness 
places no limit to the resolution. On the contrary, lower energy favors 
larger resolution. Moreover, we also show that the larger the deviation 
of the probe state from being minimum-uncertainty state, the better the 
resolution, contrary to the more intuitive idea that improved resolution 
would require probes with minimum uncertainty. These results may have 
deep implications in the development of new detection technologies free 
from quantum limits. This may also provide new insights in the still 
open vivid debate on quantum metrology limits \cite{D0,D1,D2,D3,D4,D5,D6,D7}. 

The structure of any signal-detection scheme is quite universal. 
The information about some signal $\chi$ is encoded in a system state 
by a signal-dependent transformation $U_\chi = \exp (i \chi G)$ acting 
on a probe previously prepared in a known state $| \psi \rangle$, where 
$G$ is the generator of the transformation. The transformed state 
$U_\chi | \psi \rangle$ is monitored by a measurement $M$ whose outputs 
provide an estimator $\tilde{\chi}$ of the signal $\chi$ with some 
uncertainty $\Delta \tilde{\chi}$ depending in general on $\chi$, $G$, 
$M$, $| \psi \rangle$, and the data analysis followed. 

In practical terms, $G$ is essentially an interaction Hamiltonian $H_I$ 
coupling the system probe to the external variables to be monitored. In 
an impulsive regime, the interaction Hamiltonian $H_I$ is so strong, and 
the interaction time $\tau$ so short, that the dynamics is controlled 
entirely by $H_I$. Thus the signal $\chi$ depends on the coupling between 
the system and the external variables and the interaction time.

Most approaches assume that the generator of the transformation coincides 
with the energy $H$ of the probe system in absence of signal, $G \propto 
H_I \propto H$. This is typically the case of phase shifts in harmonic 
oscillators generated by the number operator $G \propto a^\dagger a$, where 
$a$ is the complex amplitude operator. The importance in physics 
of harmonic oscillators may suggest that this is an universal link due to 
very fundamental quantum features such as uncertainty relations. The 
examples developed in this work shows that this is not so, and when 
$G \neq H$ the finiteness of energy resources may place no limit on the 
resolution.

The condition $G \neq H$ leads naturally to the so called quantum 
nondemolition strategies \cite{nd1,nd2}, in which the observable $M$, in 
absence of signal, is a constant of the motion $[H,M] =0$. Since observable 
signal-induced transformations require $[G,M] \neq 0$, quantum nondemolition 
$[H,M] = 0$ implies that $G \neq H$, which is exactly what we are looking 
for. There are two typical quantum nondemolition measurements: (i) linear 
momentum for a free particle $M= p$, $H \propto p^2$, and (ii) energy itself 
$M=H$, for example for an harmonic oscillator. Next we examine both 
possibilities from the perspective of potential ultimate quantum limits 
caused by finiteness of resources.

The performance improvement available when $G \neq H$ is at the heart of 
previous proposals of beating quantum limits by nonlinear detection schemes 
where, roughly speaking, $G \propto H^k$ \cite{NL1}. This has been recently 
confirmed experimentally \cite{NL2}.

\section{Free particle}    

\subsection{Probe state}

Let us consider that the probe system in absence of signal is the 
one-dimensional dynamics of a free particle. The signal is a shift of 
linear momentum to be inferred by a measurement of the momentum. This is 
to say $H= p^2$, $G=x$, and $M=p$, where $x$ and $p$ are suitably scaled 
dimensionless position and momentum operators with $[x,p]=i$. Following 
some previous works \cite{yo1,yo2,ZZ} we consider the probe to be prepared 
in the pure state in the momentum representation
\begin{eqnarray}
\label{ps}
& |\psi \rangle = \int_{-\infty}^\infty dp \psi( p) | p \rangle , & \nonumber \\
& \psi (p) = \langle p |\psi \rangle = 
\sqrt{\frac{\alpha 2^{1/\alpha}}{2 \gamma \Gamma (1/\alpha)}} 
\exp \left ( -  \left | p/ \gamma \right |^\alpha 
\right )  , &
\end{eqnarray}
where $| p \rangle$ are the momentum eigenstates, $\Gamma$ is the gamma 
function, and $\alpha$, $\gamma$ are real nonnegative parameters. The momentum 
statistics in absence of signal is
\begin{equation}
\label{Pp}
P(p) = \left | \psi (p) \right |^2 = \frac{\alpha 2^{1/\alpha}}{2 \gamma 
\Gamma (1/\alpha)} \exp \left ( - 2 \left | p/ \gamma \right |^\alpha \right ) .
\end{equation}
We will consider $\alpha$ to be even integers in order to avoid continuity 
problems of derivatives at $p=0$. For $\alpha = 2$ these are Gaussians, while 
for $\alpha \rightarrow \infty$ they are super Gaussians that tend to be a 
square distribution (see Fig. 1). These probes may be produced by letting 
free particles pass consecutively through rotating slits with suitable width 
and separation. 

\begin{figure}
\begin{center}
\includegraphics[width=8cm]{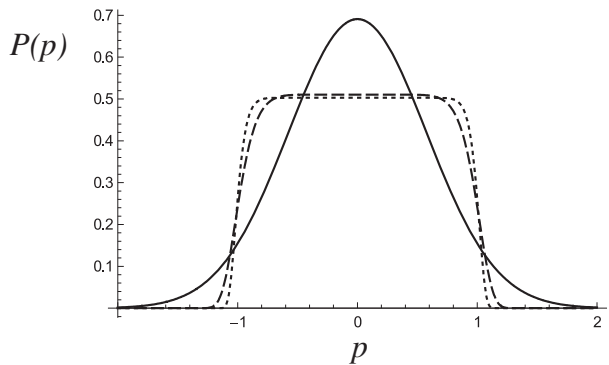}
\end{center}
\caption{Plot of the momentum statistics $P(p)$ of the initial state 
of the probe as a function of $p$ for $\alpha =2$ (solid line), 
$\alpha = 10$ (dashed line), and $\alpha= 20$ (dotted line), all with 
$\gamma$ chosen so that $\langle H \rangle =1/3$. }
\end{figure}

\subsection{Signal-induced transformation in Heisenberg and Schr\"{o}dinger 
pictures}

The momentum shift may be created by a large force acting an extremely short time, 
i. e., an impulse. This force results from an interaction Hamiltonian of the form 
$H_I = - \lambda x$, where $\lambda$ is a suitable constant. In the Heisenberg picture, 
the equations of motion are 
\begin{equation}
\dot{p} = -i[p,H+H_I]= \lambda, \qquad \dot{x} = -i[x,H+H_I] = 2 p, 
\end{equation}
with $H + H_I = p^2 - \lambda x$. After an interaction time $\tau$ the transformed 
position $\tilde{x}$ and momentum $\tilde{p}$ in terms of the original ones $x$, $p$, 
are   
\begin{equation}
\tilde{p} =\lambda \tau + p , \qquad
\tilde{x}= \lambda \tau^2 + 2 \tau p + x ,
\end{equation}
where, roughly speaking, $\tau$ is given by the time spent by the particle 
within the interaction region. In an impulsive regime the interaction $H_I$ is 
strong compared to free energy $H$, and is acting during a time interval $\tau$ 
short compared with typical evolution times of the free system. Roughly speaking, 
$\lambda \rightarrow \infty$ and $\tau \rightarrow 0$, with finite $\chi = 
\lambda \tau$. In such a regime the effect of the free term $H$ is negligible 
and the transformation after the interaction time $\tau$ reads approximately
\begin{equation}
\tilde{p} =\chi + p = U^\dagger p U, \qquad 
\tilde{x} \simeq x = U^\dagger x U, 
\end{equation}
where
\begin{equation}
\chi = \lambda \tau, \qquad
U = \exp (- i H_I \tau ) = \exp (i \chi x ) ,
\end{equation}
and the signal-dependent transformation is $U=\exp (i \chi G )$ with $G=x$. 

On the other hand, in the Schr\"{o}dinger picture the transformed probe state is
\begin{eqnarray}
| \tilde{\psi} \rangle & = & U | \psi \rangle  = \int_{-\infty}^\infty dp \psi( p) 
U | p \rangle = \int_{-\infty}^\infty dp \psi( p) | p + \chi \rangle \nonumber \\
& = & \int_{-\infty}^\infty dp \psi( p - \chi)| p \rangle ,
\end{eqnarray}
so that the momentum statistics $\tilde{P}(p)$ of the transformed probe state 
is
\begin{equation}
\label{Pp}
\tilde{P}(p) = P(p - \chi) = \frac{\alpha 2^{1/\alpha}}{2 \gamma \Gamma (1/\alpha)} 
\exp \left ( - 2 \left | \frac{p - \chi}{ \gamma} \right |^\alpha \right ) .
\end{equation}

\subsection{Examples of interaction Hamiltonian}

As a first suitable practical implementation of the desired interaction 
Hamiltonian $H_I = - \lambda x$ we may consider the interaction between 
a charged particle and a classical electric field $E$ in the dipole 
approximation with just a nonvanishing component along the $x$ axis
\begin{equation}
H_I = - q x E,   \qquad  \chi = q E \tau ,
\end{equation}
where $q$ is the electric charge. In this case the signal $\chi$ may 
represent either an electric field, the charge, or the interaction time.

As a second example we may consider an Stern-Gerlach device with the 
interaction between a magnetic moment $\mathbf{\mu}$ and an inhomogeneous 
magnetic field with just a nonvanishing component pointing in the $z$ axis 
$B_z \simeq B_0 x$ so that 
\begin{equation}
H_I = - \mu_z B_0 x,   \qquad  \chi = \mu_z B_0 \tau .
\end{equation}
In this case the signal $\chi$ may represent either the gradient of a 
magnetic field, a component of the magnetic moment, or the interaction 
time.

Within a quantum field framework, sudden momentum kicks arise from 
momentum conservation in the absorption/emission of photons during the 
interaction between the charge and the field. Thus $\chi$ represents 
the strength of the scattering of photons by the charge.

\subsection{Signal uncertainty and main results}

Signal resolution is estimated in the standard way via the Cram\'{e}r-Rao
lower bound providing a minimum for the estimator uncertainty $\Delta \tilde{\chi}$ 
\cite{FI1,FI2}
\begin{equation}
\label{DcFQ}
\left ( \Delta \tilde{\chi} \right )^2 \geq \frac{1}{N F} \geq \frac{1}{N F_Q},
\end{equation}
where $N$ is the number of repetitions of the measurement, $F$ is the Fisher 
information
\begin{equation}
F = \int_{-\infty}^\infty dp \frac{1}{P(p| \chi)} \left ( 
\frac{dP(p|\chi)}{d \chi} \right )^2 ,
\end{equation}
and $P(p| \chi)$ is the momentum statistics conditioned to the signal value 
$\chi$. $F_Q$ is the quantum Fisher information, that for pure states reads 
$F_Q = 4 ( \Delta G )^2 = 4 (\Delta x)^2$. Note that free evolution does not 
affect the statistics since momentum is preserved. This is a key feature of 
this approach as a quantum nondemolition scheme.

Since the signal-induced transformation is a momentum shift we have $P(p|\chi)=
P(p- \chi)$ and the Fisher information does not depend on $\chi$. We readily 
obtain 
\begin{equation}
\label{Fi}
F = F_Q = \frac{\alpha^2 2^{2/\alpha} \Gamma(2 - 1/\alpha)}{\gamma^2 
\Gamma (1/\alpha)} ,
\end{equation}
so that the scheme is efficient in the sense of reaching the maximum resolution 
allowed by the quantum Fisher information. The amount of energy involved in the 
measurement can be conveniently expressed by the mean value
\begin{equation}
\langle H \rangle = \left ( \Delta p \right )^2 = \frac{\gamma^2  
\Gamma(3/\alpha)}{2^{2/\alpha}
\Gamma (1/\alpha)} ,
\end{equation}
where we have taken into account that $\langle p \rangle =0$.

In order to analyze the role played by finiteness of  energy let us fix the 
value of $\alpha$, expressing $\gamma$ in terms of $\langle H \rangle$ to 
get a one-parameter family of probe states parametrized by $\langle H \rangle$. 
This allows us to get the following lower bound to the signal uncertainty
\begin{equation}
\label{Dchi}
\left ( \Delta \tilde{\chi} \right )^2 \geq \frac{ 
\langle H \rangle \Gamma^2 (1/\alpha)}{N \alpha^2 \Gamma(2 - 1/\alpha) 
\Gamma (3/\alpha)} .
\end{equation}

In Fig. 2 we have represented $N( \Delta \chi )^2 /\langle H \rangle$ 
as a function of $\alpha$, showing that the best scenario holds when 
$\alpha \gg 1$, so that the momentum statistics tends to be square. In 
such a case the following approximation holds 
\begin{equation}
\label{ap}
\left ( \Delta \tilde{\chi} \right )^2 \geq \frac{3 \langle H 
\rangle}{N \alpha}.
\end{equation}
It can be appreciated in Fig. 2 that the approximation works quite 
well even for small $\alpha$.

\begin{figure}
\begin{center}
\includegraphics[width=8cm]{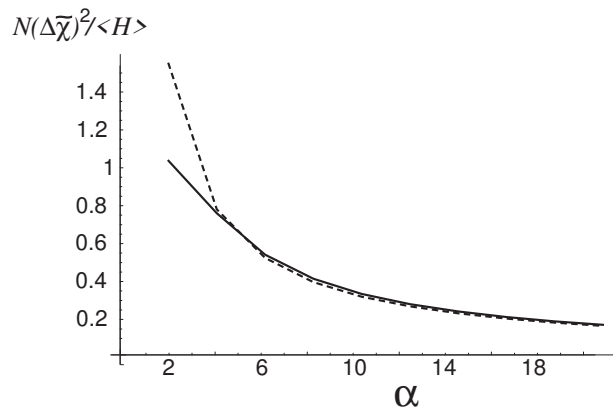}
\end{center}
\caption{Plot of $N( \Delta \tilde{\chi} )^2 /\langle H \rangle$ as a function 
of $\alpha$ (solid line) along with its approximation in Eq. (\ref{ap}) 
(dashed line), restricted to even values of $\alpha$.}
\end{figure}

This provides us with the main results of this work:

i) The lower bound on the signal uncertainty $( \Delta \tilde{\chi} )^2$ is 
proportional to $\langle H \rangle$. Thus, the lower the energy the better 
the resolution. This behavior can be clearly and simply explained. Roughly 
speaking, the inferred signal uncertainty is determined by the momentum 
uncertainty of the probe. For a free particle with $\langle p \rangle=0$ 
momentum uncertainty equals mean energy, so that lower $\langle H \rangle$ 
improves resolution.

ii) For fixed mean energy $\langle H \rangle$ the uncertainty (\ref{Dchi}) 
can be arbitrarily reduced by a proper choice of the probe state, this is 
simply by increasing $\alpha$ approaching a rectangular momentum statistics. 
This lack of limit may be ascribed to the increasingly sharp edges of $P(p)$ 
as $\alpha$ increases. 

These results are in sharp contrast with the typical situation arising in 
harmonic-oscillator detection schemes, where there is an unavoidable lower 
bound to uncertainty inversely proportional to mean energy that applies to 
every probe state (see Section 3.1 below). 

Finally note that for $\alpha \neq 2$ the lower bound in Eq. (\ref{Dchi}) 
deviates from the uncertainty $\delta \tilde{\chi}$ that may be derived 
from a much more simple analysis in terms of error propagation
\begin{equation}
(\delta \tilde{\chi})^2 = \frac{\left ( \Delta M \right )^2}{N  \left | \frac{\partial 
\langle M \rangle}{\partial \chi} \right |^2} = \frac{ \langle H \rangle}{N} .
\end{equation}
This is because for $\alpha \neq 2$ the statistics is no longer Gaussian 
and the simple mean $\tilde{\chi}=\sum_j p_j /N$ need not be an efficient 
estimator.

\subsection{Numerical simulation and uncertainty relations}

The practical approaching of the lower bound depends on the number of 
repetitions $N$ of the measurement. An estimation of the number of 
repetitions $N_B$ required is given by \cite{FI1}
\begin{equation}
N_B \simeq \frac{2}{F^2} \int_{-\infty}^\infty dp \left [ \frac{1}{P} 
\left ( \frac{d^2 P}{d p^2} \right )^2 -  \frac{1}{3 P^3} 
\left ( \frac{d P}{d p} \right )^4 \right ] - 2 ,
\end{equation}
leading in our case to  
\begin{equation}
\label{NB}
N_B \simeq \frac{2 \Gamma (2-3/\alpha) \Gamma (1/\alpha)}{\Gamma^2 
(1 - 1/\alpha)} - 2 \simeq  2 \alpha 
\end{equation}
where the last approximation holds for $\alpha \gg 1$. We have that 
the value of $N_B$ increases as resolution increases, but we get always 
accessible $N_B$ even for $\alpha$ values corresponding to almost perfect 
square momentum statistics. 

\begin{figure}
\begin{center}
\includegraphics[width=8cm]{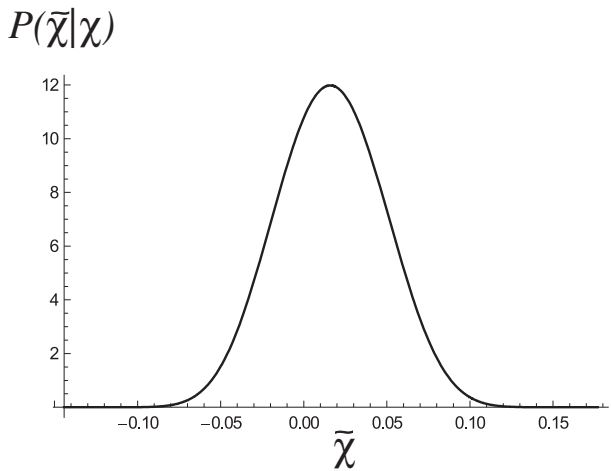}
\end{center}
\caption{Plot of a simulation of the posterior distribution $P(\tilde{\chi} | \chi )$ 
for the estimator $\tilde{\chi}$ for $\chi =0$, $\langle H \rangle = 1/3$, $N=50$ 
and $\alpha = 20$.}
\end{figure}

A simple numerical simulation confirm the result (\ref{Dchi}). In Fig. 3 we have 
represented a simulation of the posterior distribution $P(\tilde{\chi} | \chi )$ 
for the signal estimator $\tilde{\chi}$ 
\begin{equation}
P(\tilde{\chi} | \chi ) \propto \Pi_{j=1}^N P(p_j | \tilde{\chi} ) \propto 
\exp \left ( - 2 \sum_{j=1}^N \left | \frac{p_j - \tilde{\chi}}{\gamma}
\right |^\alpha \right ) ,
\end{equation}
for $\chi =0$, $\langle H \rangle = 1/3$, and $\alpha = 20$, conditioned to $N=50$ 
random outcomes $p_j$ of the measurement that have been simulated from an uniform 
distribution between $-1$ and $1$, which is very close to the actual distribution 
$P(p|\chi=0)$ for $\alpha =20$ shown in Fig. 1. It can be appreciated that 
$P(\tilde{\chi} | \chi )$ closely resembles a Gaussian distribution. In this 
example the variance of $\tilde{\chi}$ in the distribution $P(\tilde{\chi} | \chi )$
is $(\Delta \tilde{\chi} )^2 =10^{-3}$, which coincides with the lower bound in 
Eq. (\ref{Dchi}) for these parameters.  This coincidence is natural since after 
Eq. (\ref{NB}) the lower bound should be approached for $N_B \simeq 40$ repetitions.  

Finally we examine the role played by the position-momentum uncertainty relation. 
Taking into account that, for $\langle p \rangle =0$, we have $\langle H \rangle = 
(\Delta p)^2$ and $F_Q = 4 (\Delta x )^2$, we can express the signal uncertainty in 
Eq. (\ref{DcFQ}) as
\begin{equation}
\left ( \Delta \tilde{\chi} \right )^2 \geq \frac{ \langle H \rangle}{N}
\frac{1}{4 \left ( \Delta x \right )^2 \left ( \Delta p \right )^2} ,
\end{equation}
where $4 ( \Delta x )^2 ( \Delta p )^2 \geq 1$ by uncertainty relations. Note that, 
contrary to what might be assumed at first sight, the minimum signal uncertainty 
$\Delta \tilde{\chi}$ is obtained for probe estates departing from being minimum 
uncertainty states. This rather paradoxical behavior was already noticed in a different 
context in Ref. \cite{MM}.
 
\section{Harmonic oscillator}

\subsection{Finite-energy bound for $G=x$, $M=p$}

For harmonic oscillators (in suitable units) $H= p^2 + \omega^2 q^2$, so that 
neither $p$ nor $q$ are constants of the motion. Therefore, for the same 
conditions as above, i. e., $G=x$, $M=p$, we get very different conclusions. 
The mean energy, taking into account $\Delta x \Delta p \geq 1/2$, is 
\begin{eqnarray}
\label{oa}
\langle H \rangle  & = &  (\Delta p)^2 + \langle p \rangle^2 + \omega^2 (\Delta x)^2 
+ \omega^2 \langle x \rangle^2 \nonumber \\
& \geq & \omega^2 (\Delta x)^2 + \frac{1}{4 (\Delta x)^2}
\simeq \omega^2 (\Delta x)^2 ,
\end{eqnarray} 
where, since $F_Q= 4 (\Delta x)^2$ we consider $\Delta x \gg 1$, so that 
\begin{equation}
\left ( \Delta \tilde{\chi} \right )^2 \geq \frac{\omega^2}{4 N \langle H 
\rangle} .
\end{equation}
The dependence of $\Delta \tilde{\chi}$ on the mean energy $\langle H \rangle$ is 
now in the denominator, leading to an energy-depending bound that decreases for 
increasing energies. Moreover, the same bound holds unavoidably for every probe 
state with the same $\langle H \rangle$. The situation can be reverted in the 
free-particle limit $\omega \rightarrow 0$, so that the approximation in 
Eq. (\ref{oa}) no longer holds and must be instead replaced by $\langle H \rangle 
\geq 1/[4 (\Delta x)^2]$.

\subsection{Mean number shifts}

Nevertheless, within the same harmonic-oscillator context we can provide a 
more sophisticated example by considering a different choice for $G$ and 
$M$. Let us consider that the signal is encoded via number shifts. This should 
correspond to a transformation generated by the phase $G=\phi$ and monitored 
by a number measurement $M= n$. However, the proper quantum translation of 
the harmonic-oscillator phase $\phi$ finds many difficulties \cite{pp},
so there is no simple transformation of the form $U_\chi = \exp (i \chi \phi)$. 
To avoid this difficulty we can focus on the following transformation $U_\chi$ 
for small enough signals, 
\begin{equation}
U_\chi | n  \rangle \simeq | n \rangle + \sqrt{\frac{\chi}{n+1}} 
| n+1 \rangle, 
\end{equation}
where $| n \rangle$ are the eigenstates of the number operator $a^\dagger a$, 
$a^\dagger a  | n \rangle = n |n \rangle$. This transformation produces a shift 
of the mean number of photons 
\begin{equation} 
\langle n \rangle \rightarrow \langle n \rangle + \chi .
\end{equation}
If the input probe state is $| n \rangle$ and the measured observable is the 
number operator $a^\dagger a$ we get that the Fisher information and the signal 
uncertainty are:
\begin{equation}
F \simeq \frac{1}{\chi (n+1)}, \qquad \left ( \Delta \tilde{\chi} \right )^2 \geq 
\frac{\chi \left ( \langle  n \rangle + 1 \right )}{N} .
\end{equation}
We can appreciate that also within an harmonic-oscillator system we can get 
$(\Delta \tilde{\chi} )^2 \propto \langle H \rangle$. The prize to be paid is that the 
example is more abstract and less practical than the free-particle case discussed 
in Section 2. This has been introduced here merely to illustrate that results 
valid for phase shifts of harmonic oscillators do not exhausts all possibilities 
in quantum metrology.

\section{Conclusions}

We have shown that there are simple detection schemes where resolution is not 
limited by finiteness of energy resources, contrary to the standard examples 
in harmonic-oscillator systems. Moreover, we have shown that probe states 
deviating from minimum uncertainty provide better resolution. In practical 
terms this means that using schemes other than harmonic oscillators may be 
the key to improve resolution beyond currently established limits. 

\section*{Acknowledgments}

A. L. acknowledges Drs. A. Rivas and M. J. W. Hall for enlightening 
comments. This work has been supported by Project No. FIS2008-01267 
of the Spanish Direcci\'{o}n General de Investigaci\'{o}n del 
Ministerio de Ciencia e Innovaci\'{o}n, and from Project 
QUITEMAD S2009-ESP-1594 of the Consejer\'{\i}a de Educaci\'{o}n
de la Comunidad de Madrid.

\end{document}